\documentclass[preprint,showpacs,preprintnumbers,amsmath,amssymb]{revtex4}

\usepackage[dvips]{graphicx}
\usepackage{subfigure}

\begin{document}

\author{L. M. Le\'on Hilario}
\author{A. A. Aligia}
\author{A. M. Lobos}
\author{A. Bruchhausen}

\title{Polariton mediated Raman scattering in microcavities: a Green's function approach}

\pacs{71.36.+c, 78.30.Fs, 78.30.-j}

\date{\today}

\begin{abstract}
We present calculations of the intensity of polariton-mediated inelastic light
scattering in semiconductor microcavities within a Green's function framework.
In addition to reproducing the strong coupling of light and matter,
this method also enables the inclusion of damping mechanisms in a consistent way.
Our results show excellent agreement with recent Raman scattering experiments.
\end{abstract}

\affiliation{Centro At\'omico Bariloche and Instituto Balseiro, Comisi\'on Nacional de
Energ\'{\i}a At\'omica, 8400 S. C. de Bariloche, Argentina}

\maketitle


\section{Introduction}\label{sec:intro}
The strong coupling between light and exciton modes in semiconductor microcavities (MC's)
has attracted a great deal of interest in the last decade as a novel means to study and
control the interaction between radiation and matter
\cite{Skolnick_SST98,Tartakovskii_AM01,Skolnick_MC02,Fainstein_PRL95,Fainstein_PRL97,Tribe_PRB97,Fainstein_PRB98,Bruchhausen_PRB03,Bruchhausen_AIP05,Stevenson_PRB03}.
The strong coupling of photon and exciton modes in MC's leads to the formation of the \textit{polariton modes}
which are quasiparticles formed by a quantum superposition of light and matter. Polariton physics
is clearly displayed in resonant Raman scattering (RRS) experiments in which the wave length of the
incoming radiation is tuned in such a way that the energy of the outgoing radiation coincides with
that of a polariton mode, after emitting a longitudinal optical (LO) phonon, for the same conserved
two-dimensional (2D) in-plane
wave vector $k_{\parallel}$
\cite{Fainstein_PRL95,Fainstein_PRL97,Tribe_PRB97,Fainstein_PRB98,Bruchhausen_PRB03,Bruchhausen_AIP05,Stevenson_PRB03}.
Former simplified theories, which consider polaritons as the eigenstates of a two-level model,
can only qualitatively explain these experiments, and point clearly into the
relevance of including dephasing and damping effects in these
descriptions.\cite{Tribe_PRB97,Fainstein_PRB98,Bruchhausen_PRB03,Bruchhausen_AIP05}

In this work we describe the effects of including the coupling with the
electron-hole continuum, and the damping effects in RRS as mentioned in
reference \cite{Leon_JPCM07}. We consider a system where only two exciton-polariton
branches are observed \cite{Bruchhausen_PRB03}, which means that
only the $1s$ exciton plays an important role.
A survey of the formalism is presented in the Section \ref{sec:model}.
Section \ref{sec:results} contains the results and discussion, and finally
some conclusions are presented in Section \ref{sec:summary}.

\section{Model}\label{sec:model}

We use the following Hamiltonian to model the system \cite{Leon_JPCM07}
\begin{eqnarray}
H&=&E_{f}f^{\dagger}f + E_{e}e^{\dagger}e +
(Ve^{\dagger}f + H.c.) +
\sum_{p}\epsilon_{p}r^{\dagger}_{p}r_{p} +
\sum_{p}(V_{p}r^{\dagger}_{p}f + H.c.)\nonumber\\
&+& \sum_{q}\epsilon_{q}d^{\dagger}_{q}d_{q} +
\sum_{q}(V_{q}d^{\dagger}_{q}e + H.c.)+
\sum_{k}\epsilon_{k}c^{\dagger}_{k}c_{k} +
\sum_{k}(V_{k}c^{\dagger}_{k}f + H.c.)\ , \label{mod}
\end{eqnarray}
where $f^{\dagger}$ and $e^{\dagger}$ are creation operators
of the MC photon and exciton respectively. Momentum $k_{\parallel}$,
polarization and other quantum numbers have been omitted to simplify the notation.
The first three terms describe the strong coupling between the MC photon and
the exciton \cite{Tribe_PRB97,Fainstein_PRB98,Bruchhausen_PRB03,Bruchhausen_AIP05}.
The fourth and the fifth terms describe respectively a continuum of radiative modes,
 with creation operators $r_{p}^{\dagger}$,
and its coupling with the MC photon. The following two terms have a similar effect
for the exciton mode, and $d^{\dagger}_{q}$ describe a combined excitation due to
scattering with other quasiparticles (e.g. acoustic phonons). Finally, the last two
terms describe the electron-hole continuum (with $c^{\dagger}_{k}$ the creation
operator of a single excitation with relative momentum $k=k_{electron}-k_{hole}$) and its coupling with the MC photon.
This Hamiltonian can be diagonalized in the form
$H=\sum_{\nu }E_{\nu }p_{\nu }^{\dagger }p_{\nu }$,
where the boson operators $p_{\nu }^{\dagger }$ correspond to
 linear combinations of all creation operators
entering Eq. (\ref{mod}). Denoting the latter for brevity as $b_{j}^{\dagger
}$, then
$p_{\nu }^{\dagger }=\sum_{j}A_{\nu j}b_{j}^{\dagger }$. Instead of diagonalizing
the Hamitonian we will work with
retarded Green's functions $G_{jl}(\omega )=\langle \langle
b_{j};b_{l}^{\dagger }\rangle \rangle _{\omega }$. These Green's functions can be
calculated with the equation of motion method, $\omega \langle \langle b_{j};b_{l}^{\dagger }\rangle \rangle
_{\omega }=\delta _{jl}+\langle \langle [b_{j},H];b_{l}^{\dagger
}\rangle \rangle _{\omega }$.

To describe the Raman intensity we use previous models \cite{Fainstein_PRB98,Bruchhausen_PRB03},
in which $I \propto T_{i}T_{s}W$, where $T_{i}$ is the probability of conversion of an incident
photon $|f_{i}\rangle $ into a cavity eigenmode $|\nu_{i}\rangle $, $T_{s}$ has an
analogous meaning for the scattered eigenmode $|\nu_{s}\rangle $ and the
outgoing photon $|f_{s}\rangle $, and $H^{\prime }$ is the interaction
between electrons and the LO phonons. $W$ is the transition probability per unit of time,
which we calculate with the Fermi's Golden rule. Taking $T_{s}=|A_{\nu_{s}
f}|^{2}$ (the weight of photon in the scattered eigenmode) and approximating
$\langle \nu_{i} |H'|\nu_{s}\rangle \simeq A_{\nu_{s},e}$ (the excitonic
part of the scattered eigenmode), we obtain
$I=|A_{\nu_{s} e}|^{2}|A_{\nu_{s} f}|^{2}
\rho (\omega )$  where $\rho (\omega )=\sum_{j}\rho _{jj}(\omega )$ is the density
of final states, which can be calculated using the relation $\rho_{jj}(\omega)=-\frac{1}{\pi}\mathrm{Im}
G_{jj}(\omega)$, and where we have neglected the dependence of $T_{i}$ on frequency \cite{Leon_JPCM07}.
The quantities $A_{\nu j}$ can straightforwardly be calculated as
$|A_{\nu_{s}j}|^2=\rho_{jj}(\omega)/\rho(\omega)$.

When $\omega$ is chosen in such a way that the resonance condition
for the outgoing polariton is fulfilled, we obtain \cite{Leon_JPCM07}
\begin{eqnarray}
I=\frac{\rho _{ff}(\omega )\rho _{e,e}(\omega )}{\rho _{ff}(\omega )+\rho
_{e,e}(\omega )} . \label{i5}
\end{eqnarray}
The Green's functions calculated from the equation of motion are
\begin{equation}\label{green}
G_{ff}(\omega )=\frac{1}{w-E_{f}+i\delta_{f}-\frac{V^2}{w-E_{e}+i\delta_{e}}+S_{f}^{\prime}}\; ,
G_{ee}(\omega )=\frac{1}{w-E_{e}+i\delta_{e}-\frac{V^2}{w-E_{f}+i\delta_{f}}}\ ,
\end{equation}
where we have approximated the coupling with continuous modes
 $r^{\dagger}_{p}$ and $d^{\dagger}_{q}$ as purely imaginary
 contributions  $-i\delta_{f}$ and $-i\delta_{e}$ respectively.
  The electron-hole continuum states enter through the sum
  $S_{f}^{\prime }(\omega ) = \sum_{k}\frac{|V_{k}|^{2}}{\omega
+i0^{+}-\epsilon _{k}}$, and start contributing at the energy of the gap. This
corresponds to vertical transitions in which the light promotes a
valence electron with 2D wave vector $k$ to the
conduction band with the same wave vector. Using the effective-mass
aproximation we obtain \cite{Leon_JPCM07}, $S_{f}^{\prime }=R(\omega )-i(\omega -E_{XC})K \Theta (\omega -E_{XC})$ ,
where $E_{XC}$ is the bottom of the electron-hole continuum (the energy
of the semiconductor gap) and $K$ is a dimensionless parameter
that controls the magnitude of the interaction. The real part
$R(\omega )$ can be absorbed in a renormalization of the photon
energy and is unimportant in what follows. The imaginary part is a
correction to the photon width $\delta_{f}$ for energies above the bottom of
the continuum.

\section{Results and Discussion}\label{sec:results}

\begin{figure}[htbp]
       \includegraphics[width=1.0\linewidth]{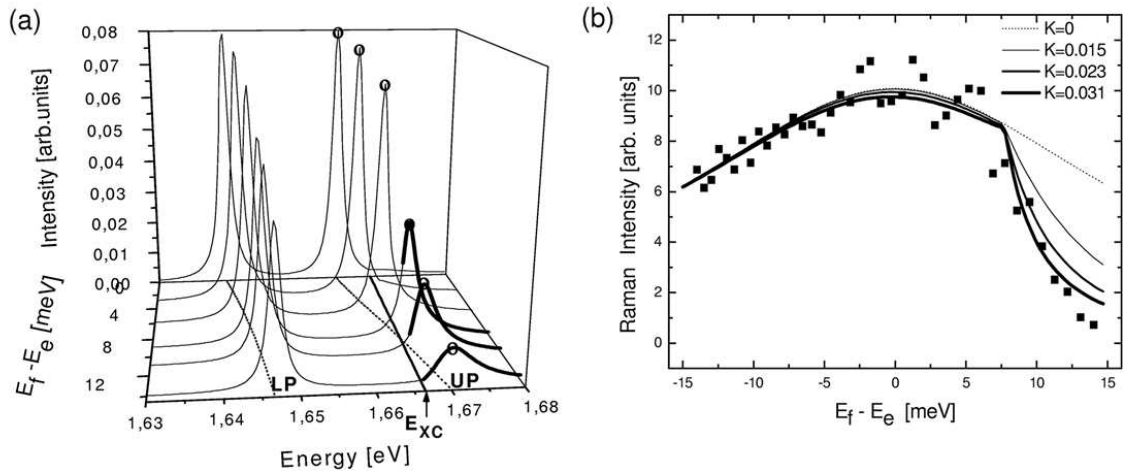}
    \caption{(a) Calculated Raman profiles as function of the Stokes shifted energy for selected fixed positive detuning ($E_{f}-E_{e}>0$), showing the LP and UP resonances. Dashed lines indicate the dispersion of the polariton branches (see text for details). (b) Raman intensity at exact resonance with the UP branch as a function of detuning. Solid squares: experimental results of Ref.~\cite{Bruchhausen_PRB03}. Solid lines: theory (Eq. \ref{i5}). Dashed line: result for a 2x2 matrix, neglecting damping effects.}
    \label{rrs1}
\end{figure}

Applying the theory outlined in the previous section, we calculated the Raman intensity for the case of a MC displaying two polariton branches, i. e. where only the $1s$ exciton mode interacts strongly with the cavity-photon mode, and a Rabi splitting of $2V\sim 19\,meV$. The parameters are chosen to be the same as those used in Ref. \cite{Leon_JPCM07}. Our results, and the comparison with the experimentally measured RRS intensities are shown in Fig.~\ref{rrs1}.
Figure~\ref{rrs1}a shows the calculated Raman intensity profiles (as function of the Stokes shifted energy) for some selected and fixed positive photon-exciton detuning ($E_{f}-E_{e}>0$). The calculated profiles show resonances at the energies corresponding to the lower (LP) and upper (UP) polariton branches, indicated by the dotted lines. The maximum of the resonance with the UP (indicated with circles in Fig.~\ref{rrs1}a), is plotted with a thick solid line in Fig.~\ref{rrs1}b as function of detuning. The experimental RRS data from Ref.\cite{Bruchhausen_PRB03} taken at exact resonance with the cavities UP is also displayed (solid squares). The agreement between experiment and theory is very good indeed. We can see that when the upper polariton (UP) mode energy increases above $E_{XC}$ entering the electron-hole continuum (thick lines in Fig.~\ref{rrs1}a), the intensity of the Raman process falls abruptly. This can be understood from  Eq.(\ref{green}): the imaginary part in the denominator increases producing a decrease in values of $\rho_{ff}(\omega)$, corresponding to processes in which light has enough energy to promote valence band electrons into the conduction band. The continuous creation and annihilation of electron-hole excitations produce a decrease of MC photon lifetime.

For completeness, we also performed analogous calculations for lower values of $K$ (thin solid lines in Fig.~\ref{rrs1}b). In the limiting case of $K \rightarrow 0$ we reobtain the expected symmetric curve (dotted line) corresponding to the case where no interaction with the electron-hole continuum is considered and damping effects are neglected.

\section{Summary}\label{sec:summary}
We have presented a Green's function approach to describe polariton mediated
Raman scattering in experiments. This method allows a consistent introduction
of damping mechanisms in polariton lifetime, being the most significant the
interaction of MC photon with the electron-hole continuum. We are able to
reproduce the main features of the experiment and the observed decrease in
the Raman efficiency when the UP polariton branch enters the continuum of electron-hole excitations.

\section*{Acknowledgments}
The authors wish to thank  A. Fainstein for useful discussions. We are partially
supported by CONICET.

\end{document}